\documentclass{emulateapj}
\newcommand{\fluxu}{$10^{-11} \, \rm erg \, cm^{-2} \, s^{-1}$}

\slugcomment{submitted to ApJ}

\shorttitle{X-Ray Outburst from M82} 
\shortauthors{Kaaret, Feng, \& Gorski}

\begin{document}

\title{A Major X-ray Outburst from an Ultraluminous X-Ray Source in M82}

\author{Philip Kaaret\altaffilmark{1}, Hua Feng\altaffilmark{2,1}, and
Mark Gorski\altaffilmark{1}}

\altaffiltext{1}{Department of Physics and Astronomy, University of
Iowa,  Van Allen Hall, Iowa City, IA 52242.}

\altaffiltext{2}{Department of Engineering Physics and Center for
Astrophysics, Tsinghua University,  Beijing 100084, China.}

%\thanks{E-mail: philip-kaaret@uiowa.edu}

\begin{abstract}

We detected a major X-ray outburst from M82 with a duration of 79 days,
an average flux of $5 \times 10^{-11} \rm \, erg \, cm^{-2} \, s^{-1}$
in the 2-10~keV band, and strong variability.  The X-ray spectrum
remained hard throughout the outburst.  We obtained a Chandra
observation during the outburst that shows that the emission arises from
the ultraluminous X-ray source X41.4+60.  This source has an unabsorbed
flux of $(5.4 \pm 0.2) \times 10^{-11} \rm \, erg \, cm^{-2} \, s^{-1}$
in the 0.3-8~keV band, equivalent to an isotropic luminosity of $8.5
\times 10^{40} \rm \, erg \, s^{-1}$.  The spectrum is adequately fitted
with an absorbed power-law with a photon index of $1.55 \pm 0.05$.  This
photon index is very similar to the value of $1.61 \pm 0.06$ measured
previously while the flux was $(2.64 \pm 0.14) \times 10^{-11} \rm \,
erg \, cm^{-2} \, s^{-1}$.  Thus, the source appears to remain in the
hard state even at the highest flux levels observed.  The X-ray spectral
and timing data available for X41.4+60 are consistent with the source
being in a luminous hard state and a black hole mass in the range of one
to a few thousand solar masses.

\end{abstract}


\section{Introduction}

The bright X-ray sources in external galaxies, known as ultraluminous
X-ray sources (ULXs), are of current interest because they represent
either an unusual state of mass accretion, not or seldom seen from
Galactic stellar-mass black hole X-ray binaries, or binary systems
containing intermediate-mass black holes \citep{Colbert99,Makishima00}. 
The ULX with the highest observed X-ray flux, CXOU J095550.2+694047 = 
X41.4+60 \citep{Kaaret01}, is located in the nearby starburst galaxy M82
at a distance of 3.63~Mpc.  If the radiation from X41.4+60 is isotropic
and from an accreting object, then the mass of the accretor must be in
excess of $500 M_{\sun}$ to avoid violating the Eddington limit. 
X41.4+60 is offset from the dynamical center of the galaxy
\citep{Kaaret01}, produces quasiperiodic oscillations (QPOs) in the
50--120~mHz range \citep{Strohmayer03,Mucciarelli06,Dewangan06,Feng07},
produces little if any radio emission \citep{Kaaret06}, and exhibits an
X-ray periodicity of 62 days \citep{Kaaret06sci,Kaaret06,Kaaret07}.  The
position of X41.4+60 is within $1\arcsec$ of the position of an infrared
source \citep{Kaaret04} identified as the super star cluster MGG 11
\citep{McCrady03} that has an extremely dense core that may have led to
numerous stellar collisions and the formation of an intermediate mass
black hole \citep{Portegies04}.

In on-going monitoring of M82 using the Proportional Counter Array (PCA)
on the Rossi X-Ray Timing Explorer (RXTE), we detected a major flare
from M82 with a peak flux corresponding to an isotropic luminosity of
$1.1 \times 10^{41} \rm \, erg \, s^{-1}$.  After the onset of this
remarkable X-ray flare, we obtained imaging X-ray observations with the
Chandra X-Ray Observatory.  We describe X-ray observations with RXTE in
\S~2, with Chandra in \S~3, and discuss the results in \S~4.

\section{RXTE observations}

We monitored M82 using the PCA on RXTE \citep{Bradt93} under programs
92098 and 93123 (PI Kaaret). As described in
\citet{Kaaret06sci,Kaaret06,Kaaret07}, we analyzed the real time data
from RXTE soon after the data became available on the RXTE ftp site
which enabled us to request a target of opportunity observation with the
Chandra X-ray Observatory \citep{Weisskopf02} after detection of the
flare.

\begin{figure*}[tb]
\centerline{\includegraphics[width=5.5in]{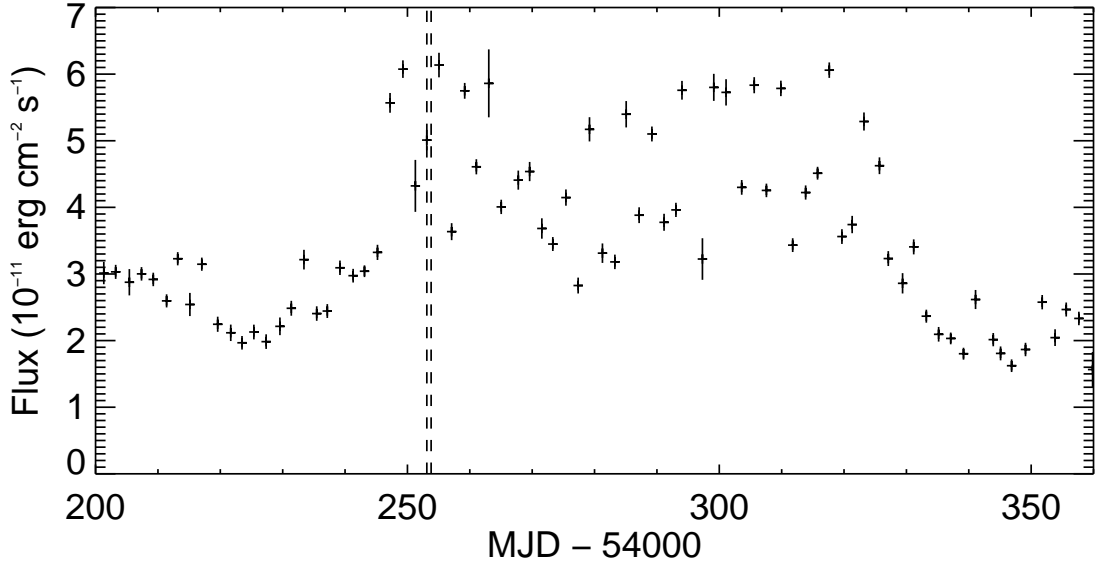}}
\caption{\label{xtelc}  Light curve of M82 obtained using the PCA on
RXTE. The plot shows the flux in the 2--10~keV band calculated for each
observation versus the observation date in MJD.  The vertical dashed
lines indicate the time of the Chandra observation.  An X-ray outburst
occurred from MJD 54247 to MJD 54326.} \end{figure*}

The RXTE results presented here were obtained with the production data
and the analysis was carried out with version 6.4.1 of the HEAsoft
package and the CALDB available in June 2008.  After the data were
retrieved, they were filtered to select good time intervals such that
Proportional Counter Unit (PCU) 2 was on, the source was more than ten
degrees above the horizon, the pointing offset from the source was less
than $0.01\arcdeg$, the satellite was more than 30~s past the South
Atlantic Anomaly (SAA), and the electron contamination was low,
specifically with an electron rate less than 0.1~c/s.  A background file
was made for the same interval as the observation with the model
pca\_bkgd\_cmfaintl7\_eMv20051128.mdl.  The tool {\tt saextrct} was used
to make both light curves and spectra for the source and background,
using only data from PCU 2.  Spectra in the 3-12~keV band were fitted
with a power-law model using the spectral fitting program XSPEC to
obtain a measurement of the flux in the 2--10~keV band.  A power-law
model with interstellar absorption with the column density fixed to $1.2
\times 10^{22} \rm \, cm^{-2}$ was used.  The average photon index for
the observations shown in Fig.~\ref{xtelc} was 2.15 with a standard
deviation of 0.21.  The average uncertainty in the photon index was
0.20, so there is no evidence for spectral variation.  Fluxes calculated
with the photon index fixed to the average value agree with those
calculated from the fits with the photon index left free.  We note that
the PCA field of view is about $1\arcdeg$ FWHM.  Thus, the PCA spectra
represent the summed spectra of all the sources in M82 and not the
spectrum of the X41.4+60 alone; the spectral index measured with the PCA
is not expected to agree with that measured for X41.4+60 alone with
Chandra.  We have reduced the PCA fluxes by a factor of 1.18 because the
PCA calibration appears to give systematically higher fluxes than
previous and other current X-ray instruments \citep{Tomsick99}.

Fig.~\ref{xtelc} shows the observed flux in the 2-10~keV band versus
time from spectral fits with the photon index fixed to 2.15.  The X-ray
flare begins at MJD 54247.2 when the flux rises above $5 \times 10^{-11}
\rm \, erg \, cm^{-2} \, s^{-1}$. The flux remains high, although highly
variable, until MJD 54325.7.  After that date, the flux returns to the
usual range for M82 of $\sim 2-3 \times 10^{-11} \rm \, erg \, cm^{-2}
\, s^{-1}$.  The flare duration was 79 days.  The peak flux was $6.1
\times 10^{-11} \rm \, erg \, cm^{-2} \, s^{-1}$, the average flux was
$4.6 \times 10^{-11} \rm \, erg \, cm^{-2} \, s^{-1}$, and the
root-mean-square (rms) variation in the flux was $9.7 \times 10^{-12}
\rm \, erg \, cm^{-2} \, s^{-1}$ indicating an rms amplitude of 21\%. 
The duration of the flare is similar to that of the flare found by
\citet{Rephaeli02}.  The level of variability is higher, but that may,
at least in part, be due to the limited coverage (only four
observations) available to  \citet{Rephaeli02}.

\begin{figure}[tb] \centerline{\includegraphics[width=2.5in]{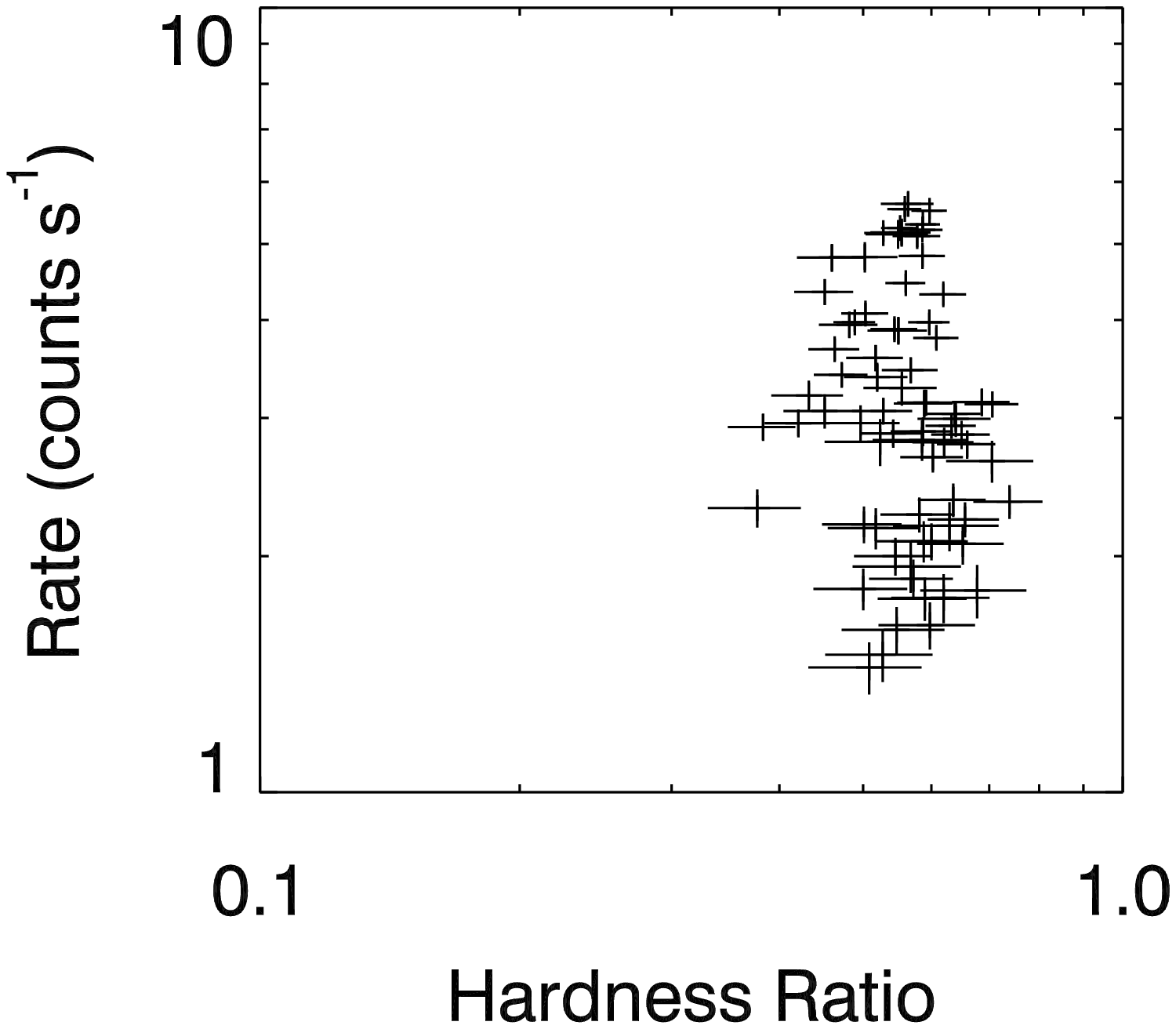}}
\caption{\label{xtehid}  Hardness-intensity diagram of M82 obtained
using the PCA on RXTE. The plot shows the count rate in the 3.6-21.6 keV
band as a function of the hardness, defined as the ratio of counts in
the 6.5-11.0 keV band to those in the 3.6-6.5~keV band.  The plot
includes those observations shown in Fig.~\ref{xtelc} for which the
hardness could be calculated to a fractional accuracy of better than
0.15.  The source remained in the hard state for the whole outburst.}
\end{figure}

Fig.~\ref{xtehid} shows a hardness-intensity diagram for M82.   We chose
energy bands similar to those used by \citet{Belloni05} in order to
facilitate comparison with previous results.  Specifically, we used the
count rate in the 3.6-21.6 keV band as the measure of intensity and the
ratio of counts in the 6.5-11.0 keV band to those in the 3.6-6.5~keV
band as the measure of hardness.  The plot includes those observations
shown in Fig.~\ref{xtelc} for which the hardness could be calculated to
a fractional accuracy of better than 0.15.  The photons recorded by the
PCA include diffuse emission and other X-ray sources in M82 that affect
the intensity and hardness.  However, the changes in intensity and
hardness are likely dominated by X41.4+60.  Fig.~\ref{xtehid} indicates
that X41.4+60 is usually in the hard state and remains in the hard state
during the whole of the outburst.  There is no significant change in
hardness as the intensity changes over a factor of about 4.

We searched for rapid variability following the procedures described in
\citet{Kaaret07}.  No statistically significant quasiperiodic
oscillations (QPOs) were found.  This may be because the new RXTE
observations are rather short, typically less than 1000~s, and have
fewer Proportional Counter Units turned on compared to previous RXTE
observations in which QPOs were detected.

\section{Observations and results}

After detection of a high X-ray flux level with the PCA, a Chandra
\citep{Weisskopf02} Target of Opportunity Observation (TOO) was
triggered.   The observation (ObsID 8190; PI Kaaret) began on  2007 June
2 at 02:41:29 UT (MJD 54253.112).  The observation was made using the
Advanced CCD Imaging Spectrometer spectroscopy array (ACIS-S) and the
High Resolution Mirror Assembly (HRMA).   We obtained 52,768~s of useful
exposure.  The ACIS-S was used in imaging mode.  As described in
\citet{Kaaret06} for our previous TOO, we attempted to reduce pile-up by
offsetting the target $-3\farcm57$ along the Y-detector coordinate in
order to image the source using a point spread function that spread the
beam over many CCD pixels and operating only the S3 chip in the 1/8
sub-array mode to reduce the frame time to 0.441~s.  We also re-analyzed
the Chandra observation (ObsID 6097) of M82 that began on 2005 February
4 at 23:34:40 UT (MJD 53405.982). 

\begin{figure*}[tb]
\centerline{\includegraphics[width=3.2in]{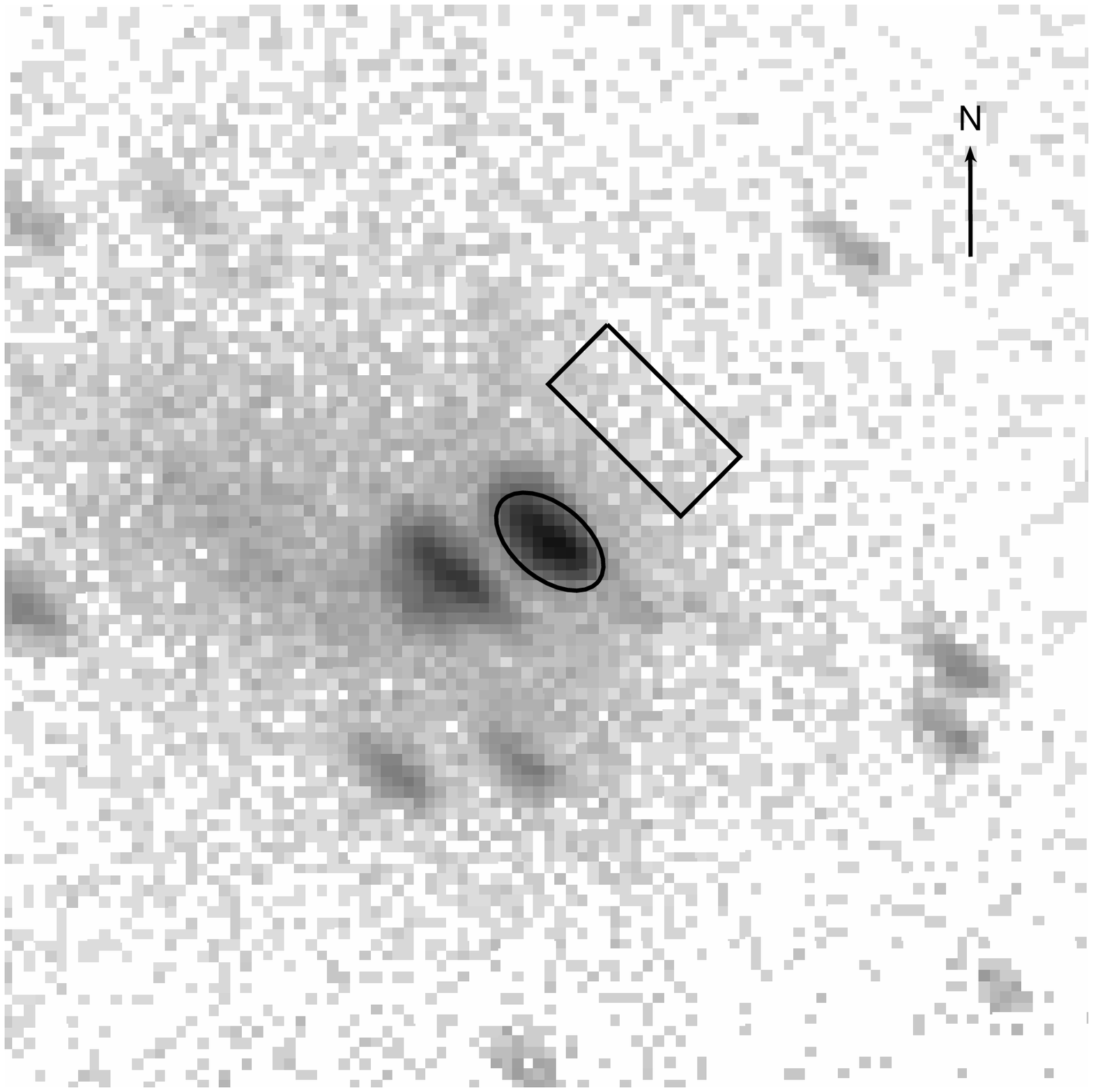} ~
            \includegraphics[width=3.2in]{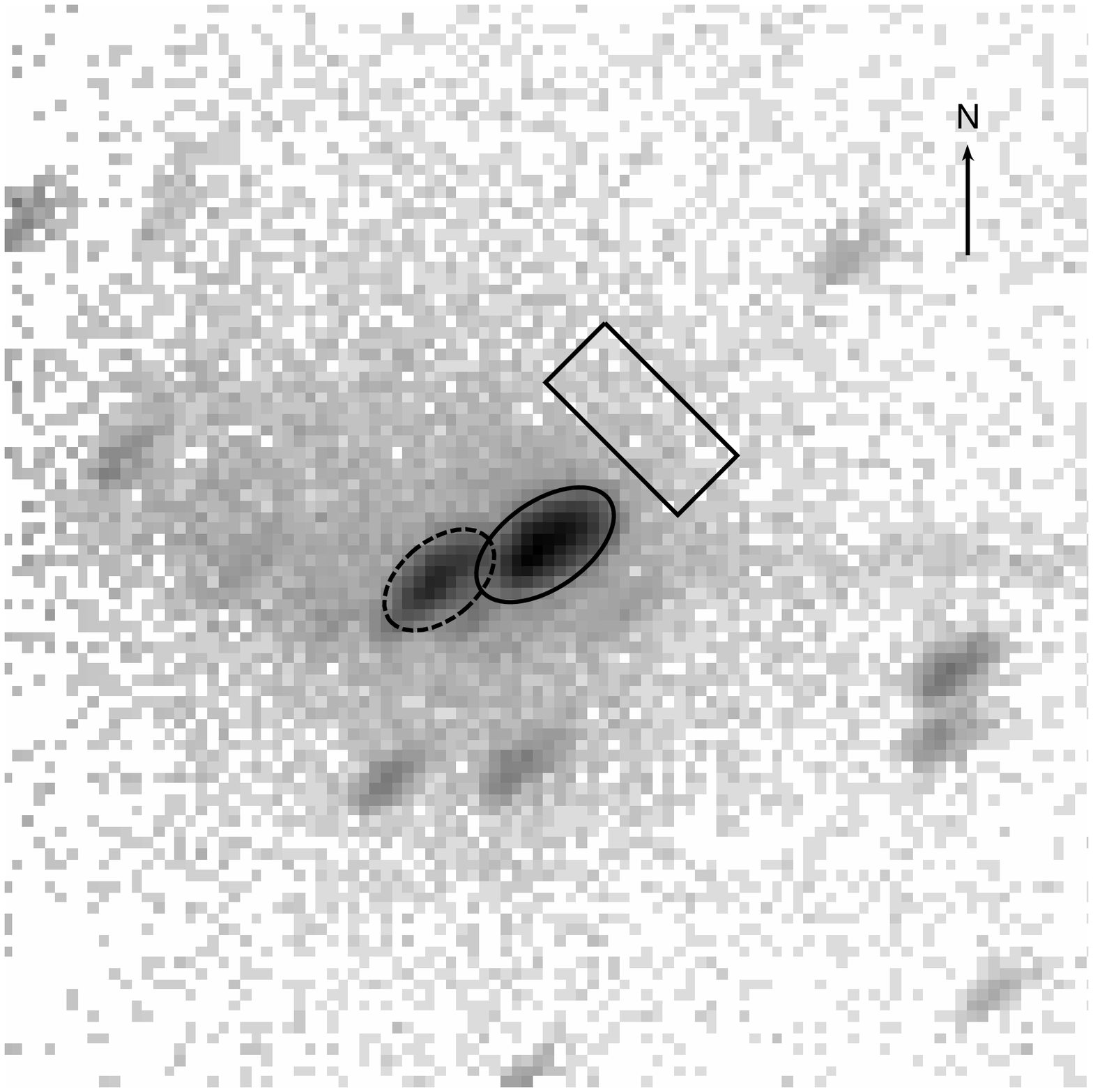}}
\caption{\label{chandra_image}  X-ray images of M82 in the 2--8~keV
band.  The panel on the left shows the Chandra observation on 2005
February 4, while the panel on the right shows the new Chandra
observation on 2007 June 2.  The intensity scale is the same in the two
images.  In both panels, the solid ellipse shows the source extraction
region, the rectangle is the background region, and the arrow points
North and has a length of $5\arcsec$.  For the 2007 observation, the
dashed ellipse shows the source region for the X-ray source X42.3+59.  
This region was excluded in the extraction of photons from X41.4+60.}
\end{figure*}

The Chandra data were subjected to standard data processing and event
screening (Ciao version 4.0.1 using CALDB version 3.4.3).  Neither
observation showed any strong background flares.  Following
\citet{Kaaret06}, we constructed images for the 2--8~keV band and used
the {\it wavdetect} tool to locate the X-ray sources. 
Fig.~\ref{chandra_image} shows the images and the source regions for
X41.4+60 defined using the 2--8~keV band.  Note that the sources appear
elongated because they are off axis.  The ellipses encompassing X41.4+60
have semi-major axes of $5.69\arcsec$ and $7.11\arcsec$, and semi-minor
axes of $3.33\arcsec$ and $2.93\arcsec$, respectively for observation
6097 and 8190.  The background region is defined by a $8.49\arcsec$ by
$3.81\arcsec$ box placed near the source, but avoiding other point
sources and the readout streak for X41.4+60.  In observation 8190, some
photons from another bright source, X42.3+59, fall into the region
defined for X41.4+60. When accumulating the spectrum for X41.4+60, we
excluded photons from the region around X42.3+59, indicated by the
dotted ellipse in Fig.~\ref{chandra_image}, in order to minimize the
contamination. Response matrices were calculated using {\it mkacisrmf}
and the spectra were fitted using {XSPEC 12} which is part of the
HEAsoft package.

\begin{deluxetable*}{lcccc}
\tabletypesize{\scriptsize}
\tablecaption{X-Ray Spectral Fits \label{spectralfits}}
\tablewidth{0pt}
\tablehead{   & \colhead{$\Gamma$} & \colhead{$N_H$} & \colhead{Flux} & \colhead{$\chi^2$/DoF}  \\
              &                    & ($10^{22} \, \rm cm^{-2}$) 
                                                     & ($10^{-11} \, \rm erg \, cm^{-2} \, s^{-1}$)
                                                                      &  }
\startdata \hline
\multicolumn{5}{l}{Observation 6097 - 2005 February 4} \\
MARX input   & 1.61                & 1.15            & 1.55           &      \\ 
MARX output  & 1.65$\pm$0.03       & 1.20$\pm$0.03   & 1.27$\pm$0.05  & 433.2/401 \\
" pileup     & 1.73$\pm$0.05       & 1.24$\pm$0.03   & 1.30$\pm$0.06  & 421.7/400 \\
Data         & 1.63$\pm$0.03       & 1.18$\pm$0.03   & 1.29$\pm$0.05  & 347.1/407 \\
" pileup     & 1.67$\pm$0.05       & 1.20$\pm$0.03   & 1.30$\pm$0.07  & 343.1/406 \\ \hline
\multicolumn{5}{l}{Observation 8190 - 2007 June 2} \\
MARX input   & 1.55                & 1.29            & 3.20           &      \\ 
MARX output  & 1.50$\pm$0.02       & 1.32$\pm$0.02   & 2.67$\pm$0.08  & 596.6/456  \\
" pileup     & 1.63$\pm$0.04       & 1.39$\pm$0.03   & 2.77$\pm$0.11  & 533.9/455  \\
Data         & 1.50$\pm$0.02       & 1.31$\pm$0.02   & 2.66$\pm$0.08  & 587.8/463  \\ 
" pileup     & 1.56$\pm$0.03       & 1.34$\pm$0.03   & 2.70$\pm$0.11  & 575.2/462  \\  \hline
\enddata 

\tablecomments{The table includes for both observations: the parameters
input to the Marx, the best fitted parameters for fits of an absorbed
power-law model and an absorbed power-law model with pile-up correction
fitted to the output of the Marx simulation and to the data.  The
columns give: $\Gamma$ - the photon index, $N_H$ - the absorption column
density, Flux -- absorbed source flux in the 0.3--10~keV band; and the
$\chi^2$ and number of degrees of freedom in the fit.} 
\end{deluxetable*}

\begin{figure}[tb] \centerline{\includegraphics[width=3.2in]{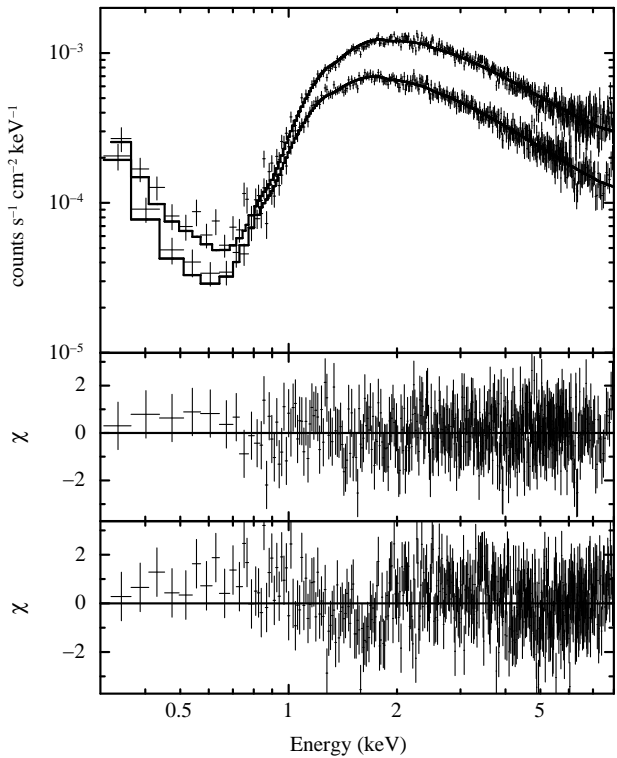}}
\caption{\label{chandra_spec}  Energy spectrum of X41.4+60.  The top
panel shows the counts~cm$^{-2}$~s$^{-1}$~keV$^{-1}$ for both
observations on the same scale.  Observation 8190 is the upper curve at
all energies.  The middle panel shows the residuals of the fit for
observation 6097.  The bottom panel shows the residuals for observation
8190.  The spectral shape at high energies is very similar between the
two observations.  The column density for observation 8190 is slightly
higher than for observation 6097.} \end{figure}

We estimated the pileup in the spectrum for each observation by summing
the counts (with no energy selection) in a 3$\times$3 pixel box around
the pixel with the maximum number of counts and then calculating the
counts per frame within the box.   Observation 6097 yielded 0.18
counts/frame and observation 8190 yielded 0.31 counts/frame.  Thus,
within the central 3$\times$3 pixel box, the fraction of piled-up events
is 5.2\% for observation 6097 and 10.3\% for observation 8190.   The
central 3$\times$3 pixel box contains 40\% and 38\% of the total events,
so the overall pile-up fraction is  about 2\% and $4$\%, respectively
for observation 6097 and 8190.

Even though the overall pile-up is low, the central regions of the image
in the second observation are strongly affected by pile-up.  We chose to
use two methods to model the pile-up.  The first was to use the ACIS
pile-up model in XSPEC version 12.4.0x.  We fitted the spectral data in
the 0.3--8~keV band with an absorbed power law with and without the
pileup model. The pileup model parameters were set to match the frame
time of the Chandra observations.  Since the observation is off-axis,
the fraction of photons in the piled-up part of the spectrum is
different from the 95\% for an on-axis point source.  We set this
fraction (the parameter psffrac) to the ratio of counts, with no energy
selection, in the 3$\times$3 pixel box around the pixel with the maximum
number of counts to the total counts in the source extraction region. 
The values are 0.40 and 0.38 for observations 6097 and 8190,
respectively.  The alpha values from the best fits are 0.33 for
observation 6097 and 0.25 for observation 8190.

To check the modeling of the pileup, the observations were simulated
using Marx version 4.3.0.  The Marx program is designed to simulate a
Chandra observation.  We set up simulations matching the off-axis
pointing and roll angle for each observation.  We followed an iterative
process to find input parameters to Marx that best matched the data: we
assumed a set of input parameter for an absorbed power-law model
(absorption column density, photon index, and normalization), ran Marx
using those parameters, fitted the output from Marx using an absorbed
power-law model, and then compared those best fitted parameters to those
obtained from the data.  Because of the high source flux, the parameters
extracted from fits to the Marx simulation without the pile-up model
differ significantly from the input parameters due to the effects of
pile-up even the off-axis pointing and the use of a 1/8${\rm th}$
sub-array.  We then adjusted the input parameters to Marx and repeated
the procedure until a good match was obtained between the fits to the
output of the simulation and the data.

The final set of parameters input to the Marx simulation, the  results
of the fits to output of that simulation, and the results of the fit to
the data are presented in Table~\ref{spectralfits}.  We take the input
to the Marx simulations as our best estimate of the true source
properties.  Comparing the Marx inputs with the fits to the data using
the power-law model with pile-up correction provide an indication of the
level of systematic uncertainty in the modeling.  Our best estimate of
the photon index is $1.61 \pm 0.06$ for observation 6097 and $1.55 \pm
0.05$ for 8190.  The absorption-corrected flux in the 0.3-8~keV band in
units of \fluxu\ is 2.64$\pm$0.14 for the first observation, and
5.4$\pm$0.2 for the second.  In the 2--10~keV band, the
absorption-corrected fluxes are 1.86$\pm$0.10 and 4.0$\pm$0.2 in the
same units.  Thus, the photon index remained constant, within the
uncertainties, while the flux changed by a factor of 2.  We note that
\citet{Berghea08} reported an unusually hard spectrum for X41.4+60 for
an observation on 1999 December 30, but this observation is strongly
affected by pile-up with a pile fraction of at least 17\% in the peak
3$\times$3 pixel box.

The spectra of some ULXs appear to have curvature at high energies,
around 5~keV and above \citep{Feng05,Stobbart06,Roberts07}, and it is of
interest to search for spectral curvature from X41.4+60.  Unfortunately,
the PCA observations around the time of the Chandra observation are too
short to make a useful measurement of the high energy extension of the
spectrum.  Also, as noted above, the PCA field of view is larger than
the angular size of M82, so the PCA spectrum would not reflect that of
X41.4+60 alone.  

We fitted the Chandra data for observation 8190 using various models
with pile-up and interstellar absorption.   We fixed the alpha parameter
to the value found in the single power-law fit as this was necessary to
prevent alpha from being driven to 1.0 (in which case all piled-up
photons are accepted as valid, negating the pile-up correction).  A
Comptonization model, {\tt comptt} in XSPEC, provided no improvement in
the fit over the power-law model and the best fit temperature was well
above 8~keV.  An exponentially cutoff power-law model, {\tt cutoffpl},
gave a cutoff energy larger than 8~keV, i.e.\ above the Chandra band.  A
multicolor disk blackbody model provides a substantially worse fit than
the simple power-law model.  A multicolor disk blackbody model in which
the temperature varies with radius as $r^{-p}$, the {\tt diskpbb},
provided an adequate fit with $p = 0.57 \pm 0.01$ and an inner disk
temperature of $6.5^{+1.3}_{-0.8}$~keV.  However, difficulties with the
physical interpretation of this model given the high apparent luminosity
and high temperature needed for X41.4+60 have already been raised by
\citet{Miyakawa08} based on spectral fitting with Suzaku.  They conclude
that the model is physically untenable and that the source appears to be
in the power-law state.  We find an even higher temperature than
\citet{Miyakawa08} which would exacerbate the problems.

\begin{figure}[tb] \centerline{\includegraphics[width=3.2in]{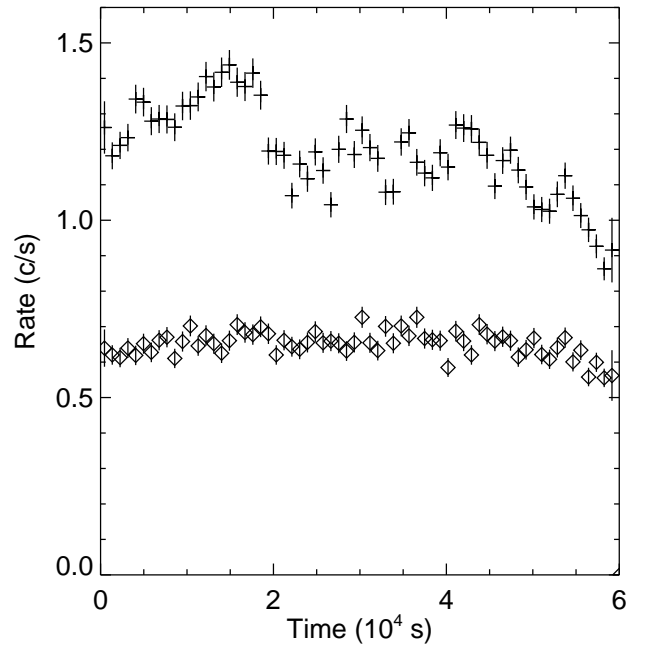}}
\caption{\label{x41lc} Light curves of X41.4+60 in the 2--8~keV band
(crosses) and the 0.3--2~keV band (diamonds) for the 2007 observation. 
The source was strongly variable in the 2--8 keV band.} \end{figure}

Fig.~\ref{x41lc} shows light curves of X41.4+60 in the 0.3--2~keV and
2--8~keV bands for the 2007 observation.  The same source region used to
extract the spectrum was used to extract the light curve.  The time bin
size is 2048 times the CCD frame time, i.e. 903.2~s.  No attempt was
made to correct for pile-up.  The source was variable on time scales of
thousands of seconds in the 2--8 keV band, but remained relatively
constant in the 0.3--2~keV band.  We divided the data into high count
rate and low count rate halves and fitted the spectra for each using an
absorbed power-law model with pile-up correction.  The absorption column
density is consistent, within errors, while the photon index for the
high count rate part, $\Gamma = 1.52 \pm 0.03$, is somewhat harder than
for the low count rate part, $\Gamma = 1.63 \pm 0.04$.  This would favor
spectral pivoting as the origin of the variability.

\section{Discussion}

The observed flux of X41.4+60 measured with Chandra on 2007 June 2 is
3.0$\times$\fluxu\ in the 2-10~keV band.   The flux from the remainder
of M82 is 1.8$\times$\fluxu.  The flux of the whole galaxy as measured
by the PCA just before the start of the Chandra observation was $(5.0
\pm 0.2)\times$\fluxu\ which is in agreement.  We adopt
1.8$\times$\fluxu\ as an estimate of the flux arising from sources other
than X41.4+60 in M82 for the PCA observations.  Then, the peak observed
flux in the 2--10~keV band from X41.4+60 during the outburst is
4.3$\times$\fluxu\ and the average flux is 2.8$\times$\fluxu.  Assuming
isotropic emission from a source located in M82 and converting these
absorbed fluxes into intrinsic fluxes using the Chandra spectrum derived
above, the peak luminosity in the 2--10~keV band is $7.6 \times 10^{40}
\rm \, erg \, s^{-1}$, the average luminosity is $4.9 \times 10^{40} \rm
\, erg \, s^{-1}$, and the total energy released in the outburst is
$\sim 3 \times 10^{47} \rm \, erg$.  Extending the energy band to the
full band, 0.3--12~keV, detected from X41.4+60 during the outburst would
increase these values by 70\%.

\citet{Remillard06} have identified three main spectral/timing states of
stellar-mass black hole X-ray binaries (BHXBs): the steep power-law
state, the thermal dominant state, and the hard state.  The Chandra
spectra show that X41.4+60 is quite hard during both observations with a
photon index near 1.6.  While it is not possible to rule out spectral
curvature above 8~keV, the observed 0.3--8~keV Chandra spectra are
inconsistent with those observed for BHXBs in the steep power-law state
or thermal dominant states.  The Chandra spectra of X41.4+60 are
adequately described by a single power-law with a photon index of
$\Gamma = 1.6$ and are consistent with identification of the source as
being in the hard state -- defined as having a spectrum dominated by a
power-law component with 80\% or more of the total flux and a photon
index $1.4 < \Gamma < 2.1$, and pronounced continuum timing noise with
an integrated rms power greater than 10\%.  We note that the photon
index at energies below any break or cutoff in the power-law component
is used by \citet{Remillard06} for the state classifications.  Thus,
identification of X41.4+60 as in the hard state is robust, even if the
spectrum shows curvature at higher energies.

The spectrum remains hard when the flux increases to the brightest
levels ever seen from the source.  The Chandra data show that the source
is variable during the second observation, but are inadequate to detect
fast timing noise.  An XMM-Newton observation made on 2004 April 21
reveals broad band timing noise \citep{Dewangan06,Mucciarelli06} which
appears to arise from X41.4+60 \citep{Feng07}.  The integrated power in
the 6-80~mHz range is near 17\%.  This value is consistent with the
magnitude of the rms power required for a source to be in the hard
state, but the frequency band is shifted relative to the 0.1-10~Hz band
used by \citet{Remillard06}.

The PCA data show that X41.4+60 remains in the hard state during the
whole of the outburst, see Fig~\ref{xtehid}.  There is no significant
change in hardness as the intensity changes over a factor of about 4. 
This lack of change in hardness is similar to the evolution of GX 339-4
in the rising phase of its 2002/3 outburst \citep{Belloni05}.  This
behavior suggests that the mass accretion rate in X41.4+60 never becomes
high enough at any point during the outburst to push the system out of
the hard state.

Stellar-mass black hole X-ray binaries have been observed to remain in
the hard state at luminosities as high as $L/L_E \sim 0.3$ for GX 339-4
\citep{Zdziarski04,Miyakawa08} and as high as $L/L_E \sim 0.1$ for XTE
J1550-564 \citep{Rodriguez03,Yuan07}, where $L$ is the source luminosity
and $L_E$ is the Eddington luminosity.  Above this luminosity, the
sources transition to a softer spectral state.  Assuming that the same
limit of $L/L_E < 0.3$ applies to more massive black holes found in the
hard state, then the 0.3--8~keV flux measured with Chandra would imply a
mass in excess of 2000~$M_{\odot}$ if the object is radiating
isotropically. Converting the peak flux observed with RXTE to a
0.3--20~keV luminosity of $1.6 \times 10^{41} \rm \, erg \, s^{-1}$
would imply a black hole mass in excess of 4000~$M_{\odot}$.  Allowing
for moderate beaming (note that a geometically thin accretion disk has a
beaming factor of 2 when viewed on axis), could reduce these values by a
factor of a few.

Using the correlation between X-ray spectral shape and $L/L_E$ found by
\citet{Shemmer08} for radio-quiet active galactic nuclei (AGN), the
photon index of $\Gamma = 1.6$ would suggest that $L/L_E \approx 0.1$. 
The highest luminosity measured with Chandra would then imply a mass of
about 7000~$M_{\odot}$.  However, mass estimation via $\Gamma$ and $L_X$
is only accurate within a factor of about 3 \citep{Shemmer08}.

Quasi-periodic oscillations have been detected from the central region
of M82 \citep{Strohmayer03,Mucciarelli06,Dewangan06} and were localized
to X41.4+60 \citep{Feng07}.  The strength and coherence of the QPOs
suggests they are of `type C' as defined for stellar-mass black hole
X-ray binaries \citep{Remillard06}.  Type C QPOs occur in the hard and
intermediate states.  Thus, the detected QPOs are consistent with the
assertion that X41.4+60 remains in the hard state.  

Estimation of a precise compact object mass using the QPO frequencies
alone is uncertain since the QPO frequencies for individual stellar-mass
black holes are known to vary by large factors, but the QPO frequencies
can be used to place bounds on the allowed range of masses.  Comparing
the maximum QPO frequency of 0.113~Hz observed from X41.4+60 with
XMM-Newton to the highest observed QPOs frequencies for GRO J1655-40,
GRS 1915+105, and XTE J1550-564 \citep{Remillard06} by scaling mass
linearly with QPO frequency, leads to an upper bound on the compact
object mass of $2.7 \times 10^4 M_{\odot}$.  If the QPO from X41.4+60 is
identified as a low frequency QPO, then a similar comparison with the
highest observed low frequency QPOs for the same stellar-mass black hole
binaries \citep{McClintock06} suggests an upper bound of $1700
M_{\odot}$. We note that the ratio of the QPO frequency to the frequency
of the break in the broad-band timing noise of X41.4+60 is consistent
with that seen for low frequency QPOs.  This, and the fact that strong
high frequency QPOs are seen only in the steep power-law state of
stellar mass black holes while low frequency QPOs are seen in the hard
state, suggests that the QPOs from X41.4+60 are analogous to low
frequency QPOs.

Better mass constraints may be obtained by combining spectral and timing
information.  We note that several authors have used XMM-Newton data to
infer the spectral shape of X41.4+60.  This is unreliable since X42.3+59
contributes significant flux is not resolved from X41.4+60 with
XMM-Newton.  Thus, all previous work suggesting the presence of a
thermal component in the spectrum or identification of the source as
being in the steep power-law or intermediate state must be re-visited
\citep{Strohmayer03,Fiorito04,Mucciarelli06,Dewangan06,Okajima06,Casella08}.
Simultaneous timing and spectral measurements of X41.4+60 will require
simultaneous observations with XMM-Newton and Chandra.

Accreting black holes of all masses are known to produce radio emission
while in the hard state.  The radio emission, X-ray emission, and black
hole mass are related via the `fundamental plane' \citep{Merloni03}.  As
noted in \citet{Kaaret01}, there was a radio transient in 1981 close to
the position of X41.4+60, but no radio emission has been detected
subsequently.  \citet{Kaaret06} place an upper limit on the radio flux
of 0.7~mJy at 8.5~GHz on 2005 February 3, just before the Chandra
observation giving a 2--10~keV luminosity of $3 \times 10^{40} \rm \,
erg \, s^{-1}$.  \citet{Paragi06} place an upper limit of 67~$\mu$Jy at
1.6~GHz on 2005 October 27, but, unfortunately, there is no
contemporaneous X-ray observation.  For upper limits on the radio
fluxes, the `fundamental plane' relation can be re-cast to place an
upper limit on the black hole mass:

$$ \textstyle
    M \lesssim 2000 M_{\odot} 
             \left( \frac{F}{100~\mu{\rm Jy}} \right)^{1.28}
             \left( \frac{L_X}{10^{40}~{\rm erg/s}} \right)^{-0.77}
             \left( \frac{d}{3.6~{\rm Mpc}} \right)^{2.56} $$

\noindent where $M$ is the black hole mass, $F$ is the flux at 5~GHz,
$L_X$ is the luminosity in the 2--10~keV band, and $d$ is the distance.
From the 2005 February data, the limit is $M \lesssim 10^4 M_{\odot}$. 
Because there is no X-ray information available for 2005 October, any
limit is less secure.  Assuming an X-ray luminosity of  $1 \times
10^{40} \rm \, erg \, s^{-1}$, would imply a limit of $M \lesssim 1200
M_{\odot}$. The scatter in the fundamental plane is significant, about a
factor of 10, so the limits from the above equation are uncertain by at
least a factor of 3.  However, simultaneous X-ray and sensitive radio
observations may provide a means to constrain the mass of X41.4+60.

There are several other known ULXs with luminosities near $10^{41} \rm
\, erg \, s^{-1}$ \citep{Gao03,Wolter06,Miniutti06}.  All of these
sources have hard spectra and are, together with X41.4+60, perhaps the
best candidates for X-ray detected intermediate-mass black holes
\citep{Roberts07}.  Unfortunately, due to the large distances to most of
these hyper luminous sources, detailed spectral information and
information on any spectral changes correlated with luminosity is
limited.  This trend toward hard spectra appears to continue down to
luminosities around $10^{40} \rm \, erg \, s^{-1}$
\citep{Feng06,Soria07,Berghea08}.  For sources with multiple
observations, particularly X-16 and X-11 in the Antennae \citep{Feng06}
and the ULX in NGC 1365 \citep{Soria07}, the sources appear to remain in
the hard state even when the flux changes by factors of several.  This
suggests that these objects also remain in the hard state.  This is
evidence against the suggestion that these sources are stellar-mass
black hole X-ray binaries that enter an ``ultraluminuous branch'' of the
very high or steep power-law state at high luminosities, since a
transition out of this state would be expected at lower luminosities.

\acknowledgements

We acknowledge partial support from Chandra grant CXC GO7-8085X and NASA
Grant NNX08AJ26G.  PK acknowledges support from a University of Iowa
Faculty Scholar Award.

{\it Facilities:} \facility{CXO (HRMA, ACIS)}, \facility{RXTE (PCA)}

%--------------

\label{lastpage}

\end{document}